\documentclass[
prc,%
10pt,%
final,%
notitlepage,%
oneside,%
twocolumn,%
nobibnotes,%
nofootinbib,
superscriptaddress,%
floatfix,%
showkeys,%
showpacs]%
{revtex4}
\usepackage{color}
\usepackage{amsfonts}
\usepackage{amsbsy}
\usepackage{mathrsfs}
\usepackage{graphicx}
\def\lsim{\mathrel{\rlap{
\lower4pt\hbox{\hskip-3pt$\sim$}}
    \raise1pt\hbox{$<$}}}     
\def\gsim{\mathrel{\rlap{
\lower4pt\hbox{\hskip-3pt$\sim$}}
    \raise1pt\hbox{$>$}}}     
\def\scr#1{\mbox{\scriptsize #1}}
\begin{document}
\title{Alternative Scenarios of Relativistic Heavy-Ion Collisions: \\
II. Particle Production}%
\author{Yu.B. Ivanov}\thanks{e-mail: Y.Ivanov@gsi.de}
\affiliation{Kurchatov Institute, 
Moscow RU-123182, Russia}
\begin{abstract}
Particle production 
in relativistic collisions of heavy nuclei 
is analyzed in a wide range of incident energies  
 2.7 GeV  $\le \sqrt{s_{NN}}\le$ 62.4 GeV. 
The analysis is performed  
within  the three-fluid model
employing three different equations of state (EoS): a purely hadronic EoS,   
an EoS with the first-order phase transition  
and that with a smooth crossover transition.
It is found that the hadronic scenario fails to reproduce 
experimental yields of antibaryons (strange and nonstrange), 
starting already from lower SPS energies, i.e. $\sqrt{s_{NN}}>$ 5 GeV. 
Moreover, at energies above the top SPS one, i.e. $\sqrt{s_{NN}}>$ 17.4 GeV, the mid-rapidity densities predicted 
by the hadronic scenario considerably exceed the available RHIC data on all species. At the same time 
the  deconfinement-transition scenarios reasonably agree (to a various extent) with all the data. 
The present analysis demonstrates certain advantage 
of the deconfinement-transition EoS's. 
However, all scenarios fail to reproduce the 
strangeness enhancement in the incident energy range near 30$A$ GeV 
(i.e. a horn anomaly in the $K^+/\pi^+$ ratio) and 
yields of $\phi$-mesons at 20$A$--40$A$ GeV.
\end{abstract}
\maketitle

\section{Introduction}

This paper continues a series of reports on simulations of relativistic heavy-ion collisions within 
different scenarios \cite{Ivanov:2012bh,3FD-2012-stopping}. 
These simulations were performed within a model of the three-fluid 
dynamics (3FD) \cite{3FD} employing three different equations of state (EoS): a purely hadronic EoS   
\cite{gasEOS} (hadr. EoS), which was used in the major part of the 3FD simulations so far 
\cite{3FD,3FD-GSI07,3FDflow,3FDpt,3FDv2}, and two versions of EoS involving the deconfinement 
 transition \cite{Toneev06}. These two versions are an EoS with the first-order phase transition
and that with a smooth crossover transition. 
Details of these calculations are described in the first paper of this series 
\cite{3FD-2012-stopping} dedicated to analysis of the baryon stopping. 


In this paper I report results on particle production and rapidity distributions of these particles in 
relativistic heavy-ion collisions in the energy range from 2.7 GeV 
to 62.4 GeV\footnote{Results for the top 
calculated energy of 
62.4 GeV should be taken with care,  
because they are not quite accurate, as an accurate computation requires 
unreasonably high memory and CPU time. 
} 
in terms of the center-of-mass incident energy ($\sqrt{s_{NN}}$). This domain covers 
the energy range of the RHIC (Relativistic Heavy-Ion Collider) 
beam-energy scan and SPS (Super Proton Synchrotron) low-energy-scan programs, as well as 
energies of the future FAIR (Facility for Antiproton and Ion Research) and 
NICA (Nuclotron-based Ion Collider Facility) facilities and the AGS 
(Alternating Gradient Synchrotron) at BNL (Brookhaven National Laboratory). 
As demonstrated in the first papers of this series 
\cite{Ivanov:2012bh,3FD-2012-stopping}, 
within the considered here first-order-transition and crossover scenarios
the  deconfinement transition takes place in the region of top-AGS--low-SPS 
incident energies. The experimental baryon stopping also indicates 
certain signs of a  deconfinement transition. Therefore, in this paper the attention is 
primaraly focused on this 
incident energy range. It should be mentioned that available data on particle production 
in the AGS-SPS energy range 
have already been analyzed within various models  
\cite{3FD,3FD-GSI07,Bratk09,Bleicher08,Bleicher09,Bratk04,WBCS03,Larionov07,Larionov05,Hama:2004rr}. 
The aim of this paper is a comparative analysis of these data within  
different assumptions on the EoS.

Closely related to the particle production, the hadron yield ratios are also discussed. 
The most intriguing issue among them is a horn anomaly in 
the $K^+/\pi^+$ ratio \cite{Gazdzicki:1998vd,Alt:2007aa} that still has no explanation 
within any dynamical model. Unfortunately, the 3FD model is not an exception from this list.

{
\section{3FD Model}\label{Model}
}

{
The 3FD model \cite{3FD} is a
straightforward extension of the 2-fluid model with radiation of
direct pions \cite{MRS88,gsi94,MRS91} and (2+1)-fluid model
\cite{Kat93,Brac97}. The above models were extend in such a
way that the created baryon-free fluid (which is called a
``fireball'' fluid, following the Frankfurt group) is treated
on equal footing with the baryon-rich ones. 
A certain formation time $\tau$ is allowed for the fireball fluid, during
which the matter of the fluid propagates without interactions. 
The formation time 
is  associated with a finite time of string formation. It is similarly 
incorporated in kinetic transport models such as UrQMD \cite{Bass98} 
and HSD \cite{Cassing99}. 
}

{
Unlike the conventional hydrodynamics, where local instantaneous
stopping of projectile and target matter is assumed, a specific
feature of the 3FD is a finite stopping
power resulting in a counter-streaming regime of leading
baryon-rich matter. The basic idea of a 3-fluid approximation 
\cite{MRS88,I87} is that at each space-time
point  a generally nonequilibrium 
distribution function of baryon-rich
matter can be represented as a sum of two
distinct contributions
initially associated with constituent nucleons of the projectile
(p) and target (t) nuclei. In addition, newly produced particles,
populating the mid-rapidity region, are associated with a fireball
(f) fluid.
Therefore, the 3-fluid approximation is a minimal way to 
simulate the finite stopping power at high incident energies.
Each of these fluids is governed by conventional hydrodynamic equations
which contain interaction terms in their right-hand sides. 
These interaction terms describe mutual friction of the fluids and 
production of the fireball fluid.
The friction between fluids was fitted to reproduce
the stopping power observed in proton rapidity distributions for each EoS, 
as it is described in  Ref. \cite{3FD-2012-stopping} in detail.
}

{
A conventional way of applying the fluid dynamics to heavy-ion collisions at
RHIC and LHC energies is to prepare the initial state for the hydrodynamics by 
means of various kinetic codes \cite{Bleicher08,Bleicher09,Hama:2004rr,Nonaka:2012qw}.
Contrary to these approaches, the 3FD model treats the collision process from the 
very beginning, i.e. the stage of cold nuclei, up to  freeze-out within  
the fluid dynamics. 
The conventional hydrodynamical model of Refs. 
\cite{Merdeev:2011bz,Mishustin:2010sd} does such simulations in a very 
similar way but without taking into account incomplete stopping of 
colliding nuclei at the initial stage of the reaction. Therefore, such kind of 
simulations are justified only at moderately high energies.  
}

{
Freeze-out  is performed accordingly to the procedure described in Ref. \cite{3FD} 
and in more detail in Refs. \cite{71,74}. This is a modified Milekhin version of the 
freeze-out that possesses exact conservation of the energy, momentum and baryon number. 
Contrary to the conventional Cooper--Frye  approach \cite{Cooper}, the modified Milekhin 
method has no problem associated with negative contributions to particle spectra. 
The freeze-out criterion is based on a phenomenologically determined   
freeze-out energy density $\varepsilon_{\scr frz}$. 
This method of freeze-out can be called dynamical, since the
freeze-out process here is integrated into fluid dynamics through
hydrodynamic equations. 
This kind of freeze-out is similar to the model of ``continuous
emission'' proposed in Ref. \cite{Sinyukov02}. There the particle emission
occurs from a surface layer of the mean-free-path width. In the 3FD case the
physical pattern is similar, only the mean free path is shrunk to zero.
To the moment of the freeze-out the matter is already in the hadronic phase in the 
case of the 2-phase EoS, while for the crossover EoS this is not so. 
However, this is not a problem because, any case,
the thermodynamic quantities of the frozen-out matter are recalculated from the 
in-matter EoS, with which the hydrodynamic calculation runs, to the hadronic 
gas EoS. This is done because  a part of the energy
is still accumulated in collective mean fields at the freeze-out instant. This mean-field
energy should be released before calculating observables.
Otherwise, the energy conservation would be violated. 
}

{
The baryon stopping turns out to be only 
moderately sensitive to the freeze-out energy density. 
The freeze-out energy density $\varepsilon_{\scr frz}=$ 0.4 fm/fm$^3$ 
was chosen mostly on the condition of the best reproduction 
of secondary particles yields. 
}

{
 In the 3FD 
model \cite{3FD}, particles are not isotopically distinguished, i.e. the model deals with nucleons, pions, etc. 
rather than with protons, neutrons, $\pi^+$, $\pi^-$, $\pi^0$, etc. In fact, it is not a 
problem to formulate the model for isotopically distinguished species. 
For that it is necessary to introduce into the EoS an additional isotopic chemical potential 
associated with the electric-charge conservation, and an additional continuity equation 
for each fluid that controls this electric-charge conservation. 
The problem comes out at the stage of numerical solution of this set of equations. 
This is the curse of dimensionality. Treatment of a 3-dimensional EoS (temperature and two 
chemical potentials) together with the (3+1)-space-time evolution
is too complicated for computers available at present. 
Thus, the disregard of the isospin is a forced approximation. 
}

{
Presently it is unconventional to apply hydrodynamics to such low incident energies as 2$A$ GeV.  
However, 20--30 years ago the hydrodynamics was a conventional tool for 
analysis of the heavy-ion collisions at incident energies around 1$A$ GeV, see e.g. Refs. 
\cite{MRS88,gsi94,Stocker,Clare}. All the arguments put forward in favor of the 
hydrodynamic approach in those references still hold true. In particular, the popular 
now blast-wave model \cite{Bondorf78,Siemens79} was originally developed for these 
low energies. 
}
\\

\section{Rapidity Distributions of Produced Particles}
\label{rapidity distributions}

The rapidity distributions of produced particles were calculated at fixed impact parameters ($b$). 
Correspondence between experimental centrality, i.e the fraction of the total reaction 
cross section related to a data set,  
and a mean value of  the impact parameter was taken from the paper \cite{Alt:2003ab}
in case of NA49 data. 
For Au+Au collisions it was approximately estimated proceeding from geometrical considerations. 
Feed-down from weak decays of strange particles into non-strange hadron yields was disregarded
in accordance with measurement procedure of the NA49 collaboration. 
At the AGS energies this contribution of weak decays is just negligible. 
At RHIC energies (Au+Au collisions at beam-energy-scan energies), the contribution of weak decays  
into non-strange hadron yields was taken into account similarly to that done in data 
of the STAR and PHENIX  collaborations.

\begin{figure*}[htb]
\includegraphics[width=18.0cm]{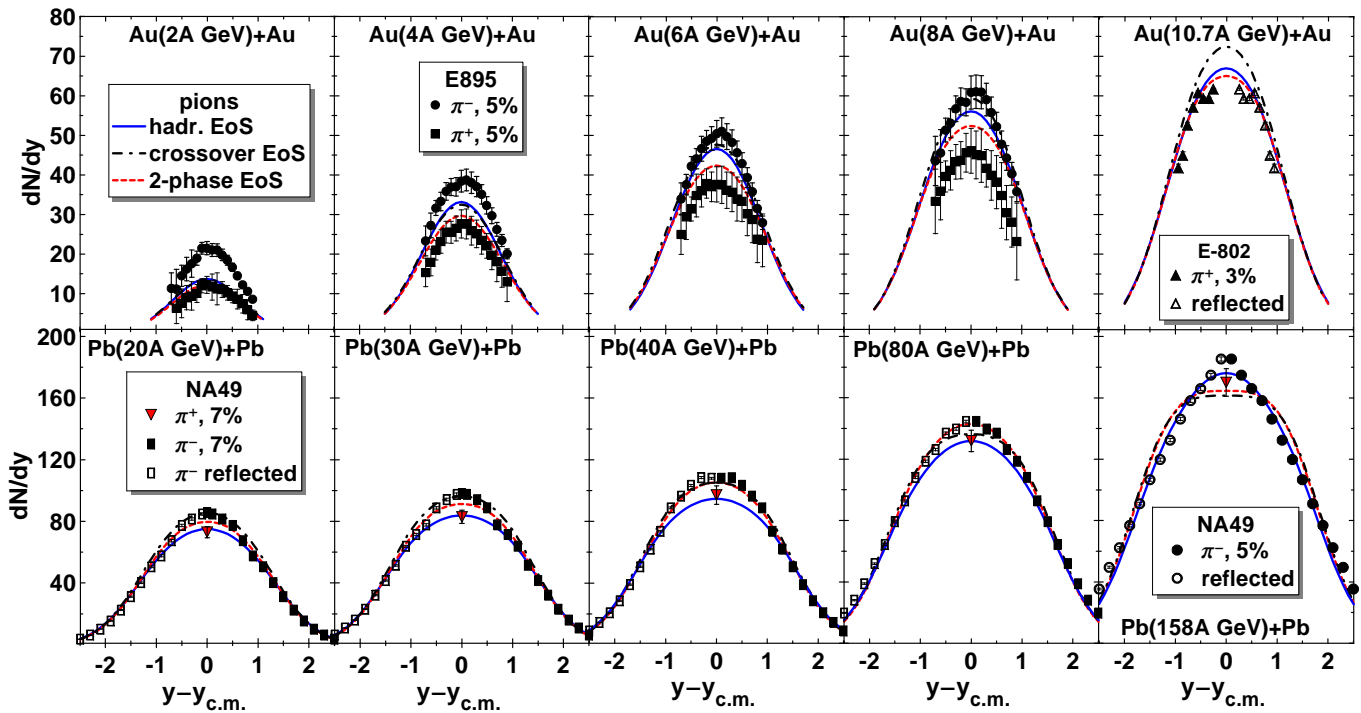}
 \caption{(Color online)
Rapidity spectra of pions
from central 
collisions of Au+Au ($b=$ 2 fm) and Pb+Pb ($b=$ 2.4 fm).   
Experimental data are taken for central Au+Au
collisions at AGS energies \cite{Klay:2003zf} (at 2$A$--8$A$ GeV, centrality 5\%) and 
\cite{Ahle:1999jm} (at 10.7$A$ GeV, centrality 3\%), and Pb+Pb collisions at SPS energies
\cite{Afanasiev:2002mx,Alt:2007aa}  
with centrality 7\% at  20$A$--80$A$ GeV  and centrality 5\% at 158$A$ GeV. 
} 
\label{fig_pi}
\end{figure*}

In Fig. \ref{fig_pi},  comparison with available experimental data on rapidity distributions 
of pions from central collisions is presented. 
As particles are not isotopically distinguished in the 3FD 
model \cite{3FD}, only the total number of pions $N_{\pi}$ is 
calculated in the model. In Fig. \ref{fig_pi}, rapidity distribution of $N_{\pi}/3$
pions is displayed which, strictly speaking, should be compared with the 
sum of $\pi^+$, $\pi^-$ and $\pi^0$ distributions divided by 3. 
In view of the fact that the experimental $\pi^-$ distributions  are always situated above 
the $\pi^+$ ones, a ``good agreement'' with the model means that calculated 
spectrum  of $1/3$ of all pions turns out in between experimental 
$\pi^+$ and $\pi^-$ distributions. 
As seen from Fig. \ref{fig_pi}, this is always the case at AGS energies for all three EoS's. 
At the SPS energies differences between predictions of different scenarios 
and experimental data do not 
exceed 10\%, which is quite acceptable in view of uncertainties of the model 
(such as a choice of the impact parameter, accuracy of the multi-fluid approximation, 
phenomenological freeze-out procedure, etc.). 
Therefore, we can approximately conclude that the 
pion rapidity distributions are reasonably well reproduced by all of the considered 
EoS's in the region of AGS--SPS incident energies.

A comment concerning the hadronic EoS is in order here. 
As it was mentioned in Refs. \cite{3FD-2012-stopping,3FD}, we proceeded from 
a principle of fair treatment of any EoS. It means that any possible uncertainties
in the parameters are treated in favor of the EOS. Therefore, for the hadronic EoS the parameters of 
the friction enhancement and the formation time of the fireball fluid 
\cite{3FD-2012-stopping,3FD} were chosen such that 
the baryon stopping and the mid-rapidity pion density were reproduced at the top SPS 
energy of 158$A$ GeV. Thus, it is not surprising that the hadronic-EoS result 
is the best at 158$A$ GeV, see Fig. \ref{fig_pi}. 
However, even with this fit the pion distributions at midrapidity, i.e. $(\pi^+ + \pi^-)/2$,  
turn out to be somewhat underestimated at lower SPS energies, as seen in Fig. \ref{fig_pi}, 
and strongly overestimated at higher (RHIC) energies, as it will be seen below.

\begin{figure*}[phtb]
\includegraphics[width=12.0cm]{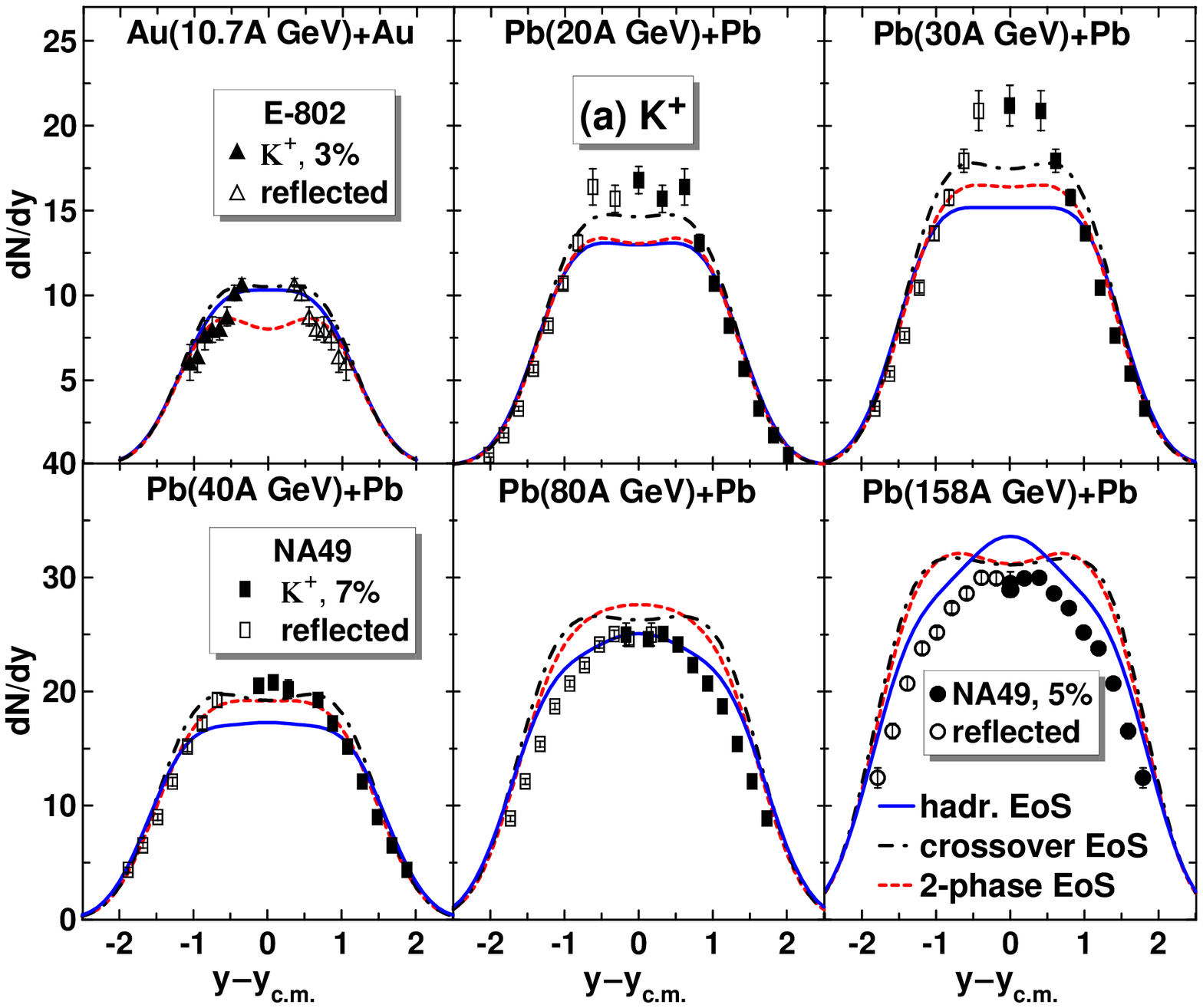}
\\
\includegraphics[width=12.0cm]{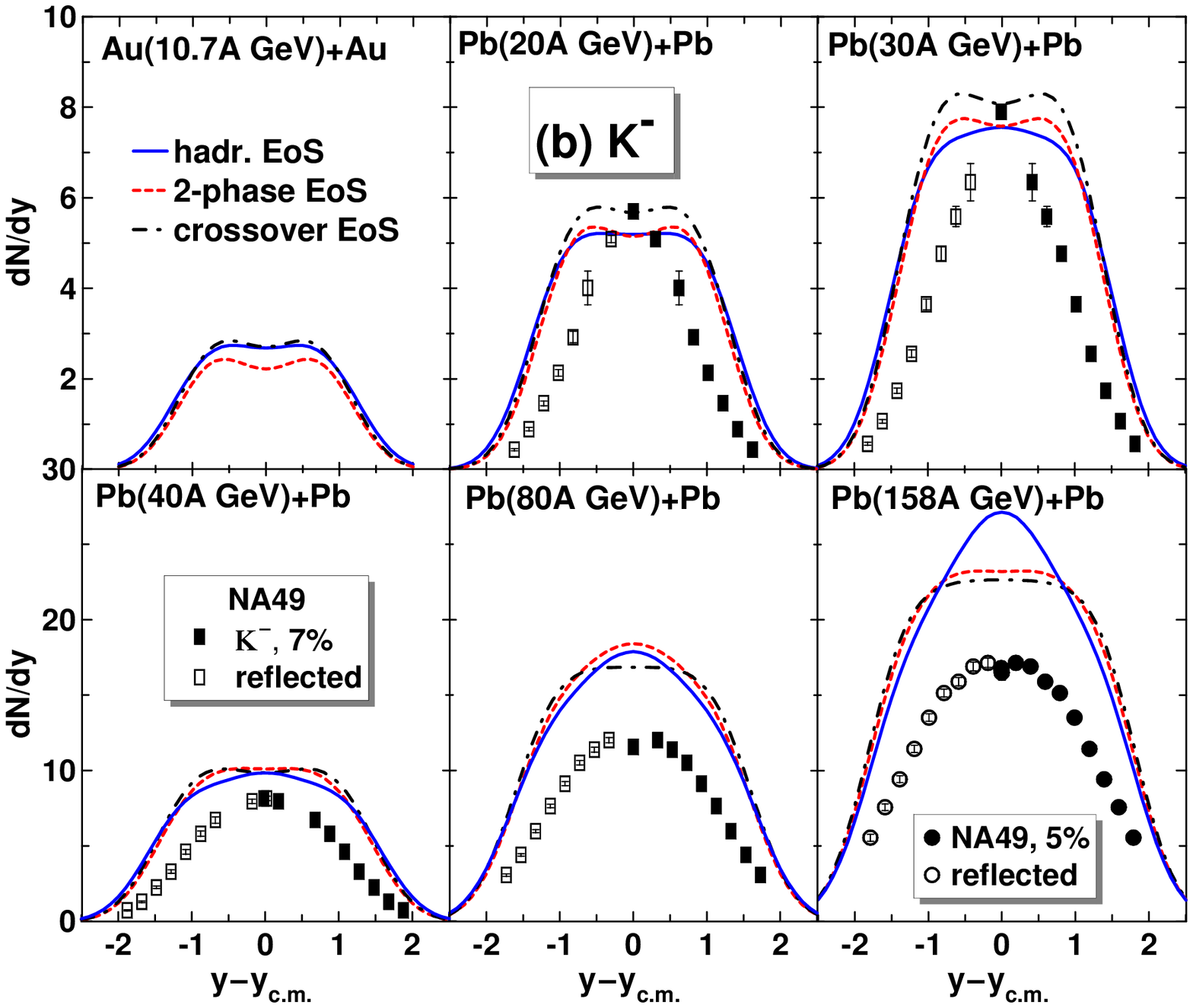}
 \caption{(Color online)
Rapidity spectra of positive [(a) block of panels] and negative [(b) block panels] kaons
from central 
collisions of Au+Au ($b=$ 2 fm) and Pb+Pb ($b=$ 2.4 fm).   
Experimental data are for central Au+Au
collisions at AGS energy of 10.7$A$ GeV (centrality 3\%)  
\cite{Ahle:1999jm} and Pb+Pb collisions
\cite{Afanasiev:2002mx,Alt:2007aa}  at SPS energies   
with centrality 7\% at  20$A$--80$A$ GeV  and centrality 5\% at 158$A$ GeV. 
} 
\label{fig_K}
\end{figure*}
\begin{figure*}[pthb]
\vspace*{-3mm}
\includegraphics[width=13.0cm]{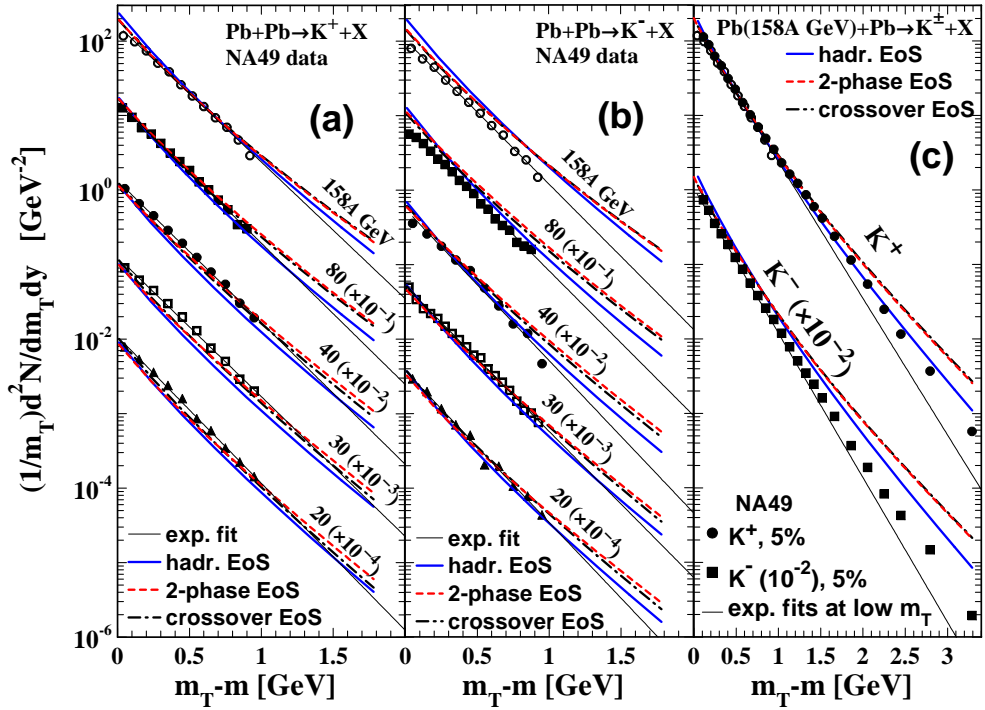}
 \caption{(Color online)
Transverse mass spectra at mid-rapidity of positive (a) and negative (b) kaons
from central collisions  Pb+Pb ($b=$ 2.4 fm) at various incident energies and those in the 
extended transverse-mass range at 158$A$ GeV (c).   
Experimental data are taken from Refs. 
\cite{Afanasiev:2002mx,Alt:2007aa} and for high $p_T$ momenta, from 
Ref. \cite{Alt:2007cd}. 
} 
\label{fig_K-mt}
\end{figure*}
%
%
\begin{figure*}[pbht]
\includegraphics[width=12.0cm]{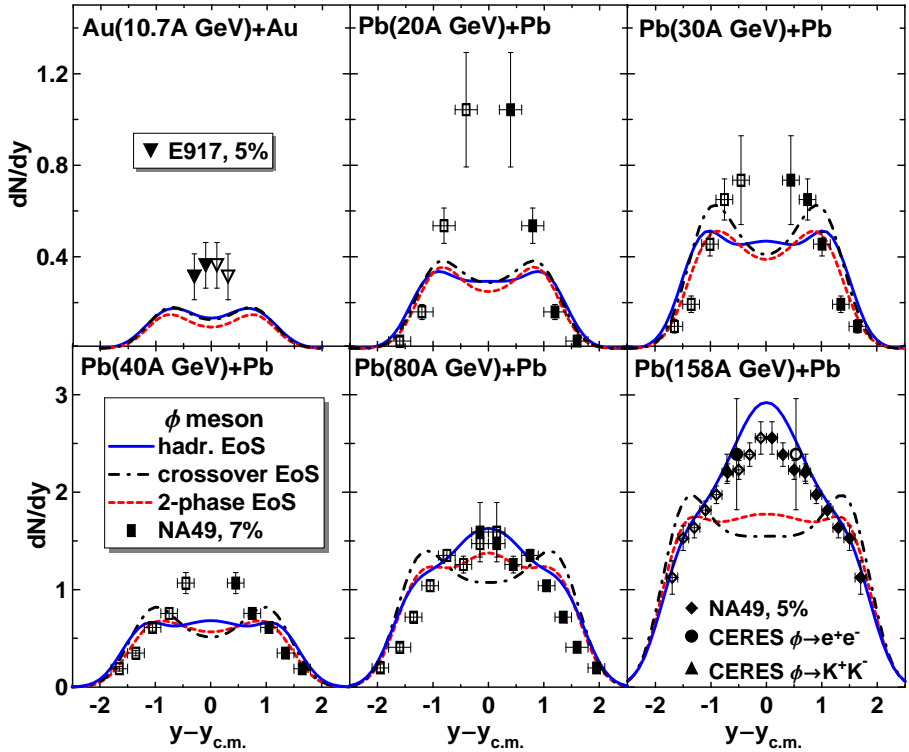}
 \caption{(Color online)
Rapidity spectra of $\phi$ mesons 
from central 
collisions of Au+Au ($b=$ 2 fm) and Pb+Pb ($b=$ 2.4 fm).   
{
Experimental data for central  Pb+Pb collisions
at SPS energies  
with centrality 7\% at  20$A$--80$A$ GeV  and centrality 5\% at 158$A$ GeV
are taken Ref. \cite{Alt:2008iv} and for central  Au+Au collisions 
at  11.7$A$ GeV/c  and centrality 5\%, from Ref. \cite{Back:2003rw}. 
}
} 
\label{fig_phi}
\end{figure*}
%
%
\begin{figure*}[phtb]
\includegraphics[width=12.0cm]{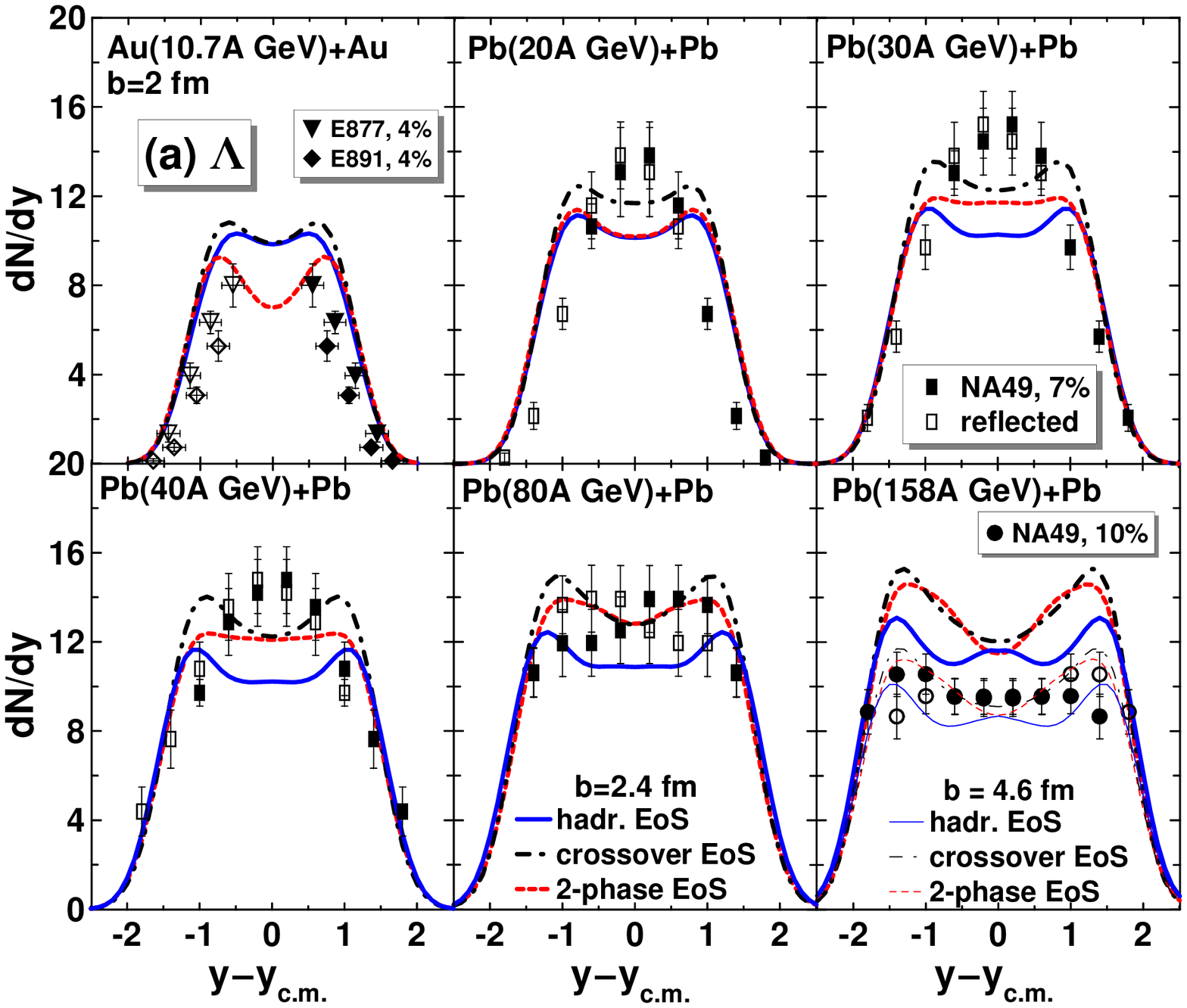}
\\
\includegraphics[width=12.0cm]{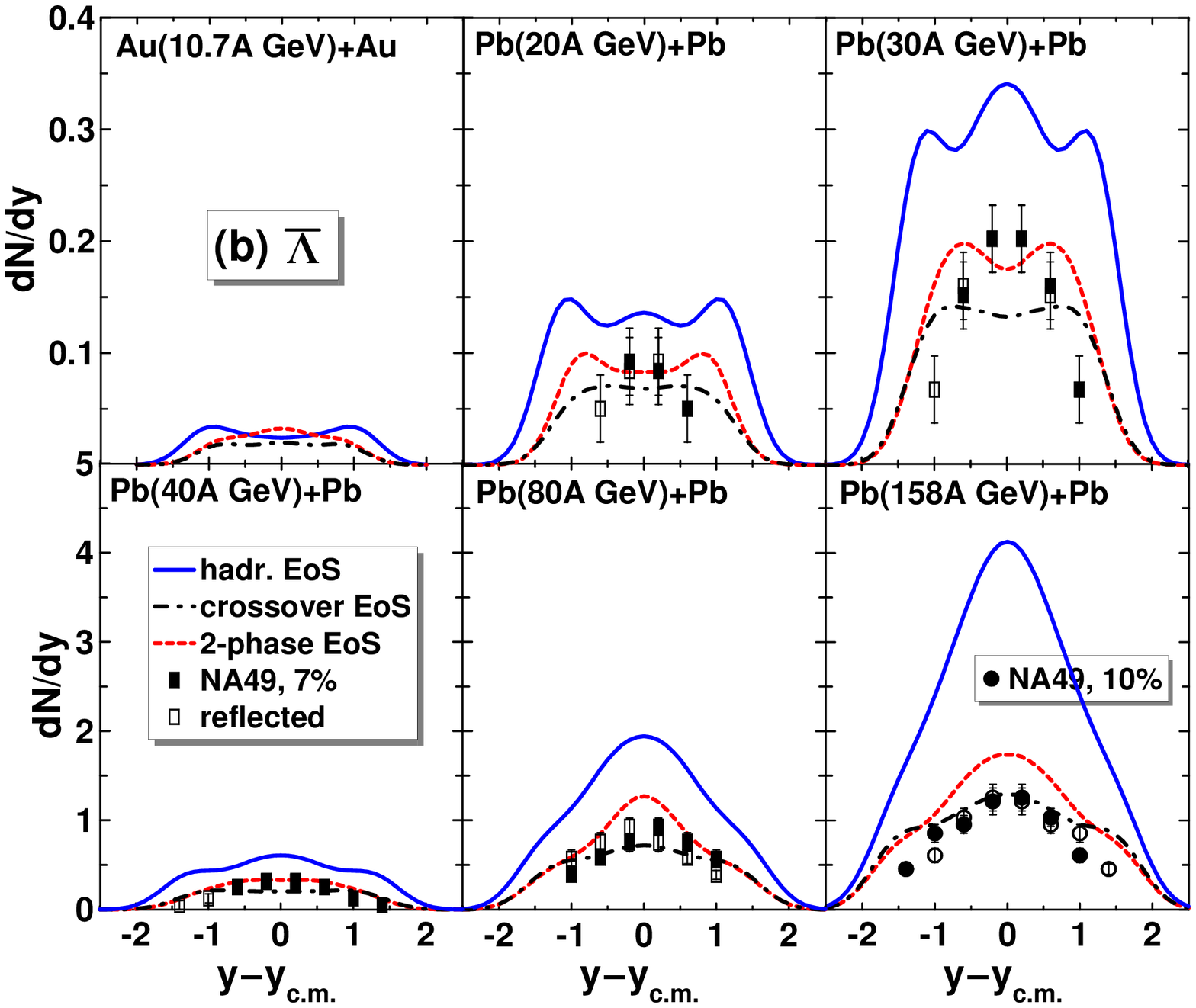}
 \caption{(Color online)
Rapidity spectra of $\Lambda$ [(a) block of panels] and $\bar{\Lambda}$  [(b) block of panels] hyperons
from central 
collisions of Au+Au ($b=$ 2 fm) and Pb+Pb ($b=$ 2.4 fm).   
{
Results for Pb+Pb collisions
with $b=$ 4.6 fm are also displayed for $\Lambda$'s at 158$A$ GeV.  
Experimental data for central  Pb+Pb collisions
at SPS energies 
with centrality 7\% at  20$A$--80$A$ GeV  and centrality 10\% at 158$A$ GeV 
are taken from Ref. \cite{Alt:2008qm}, and for central  Au+Au collisions 
at  11.5$A$ GeV/c  and centrality 4\%, from Ref. \cite{Barrette:2000cb}.
}
} 
\label{fig_L}
\end{figure*}
\begin{figure*}[phtb]
\includegraphics[width=12.0cm]{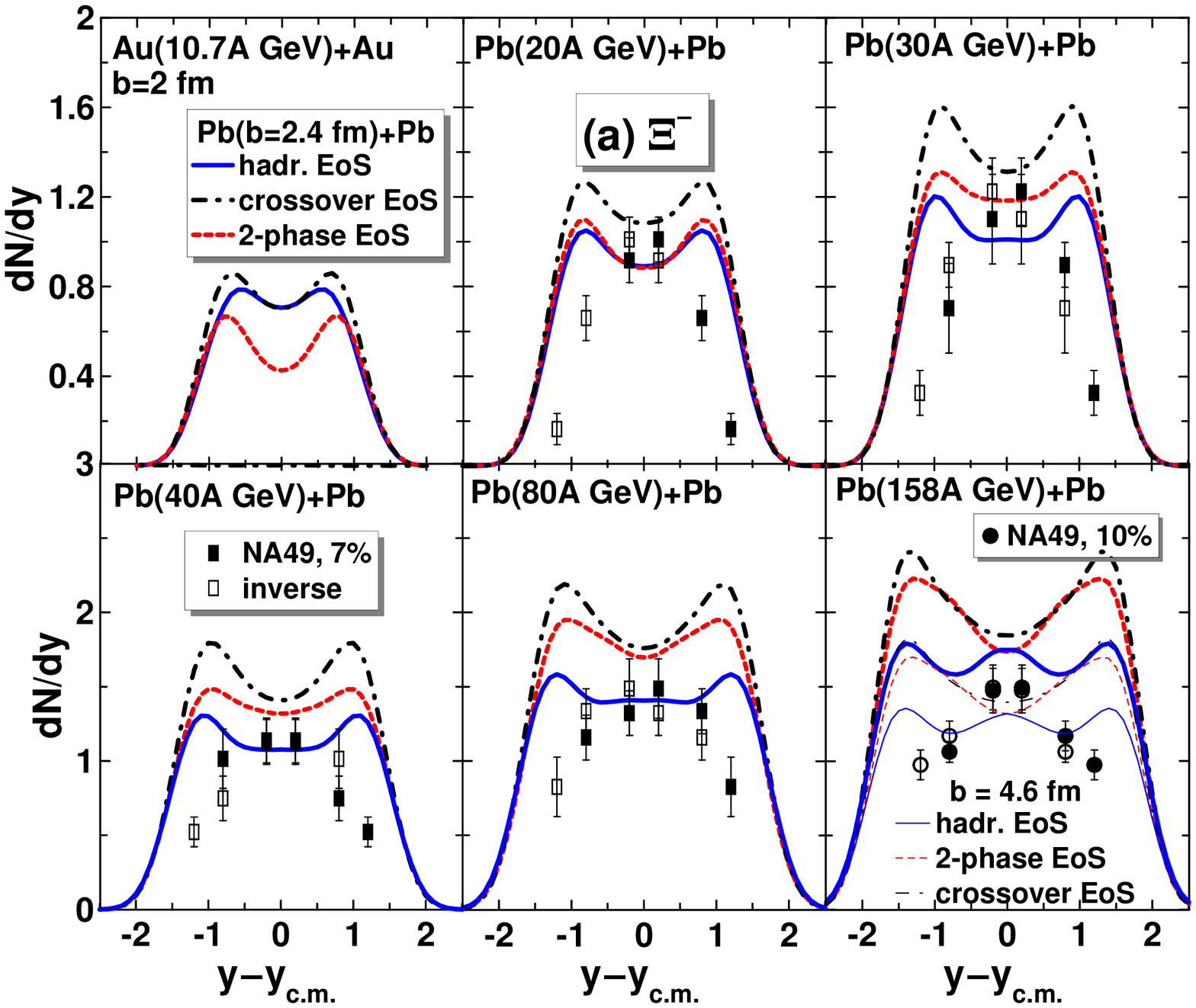}
\\
\includegraphics[width=12.0cm]{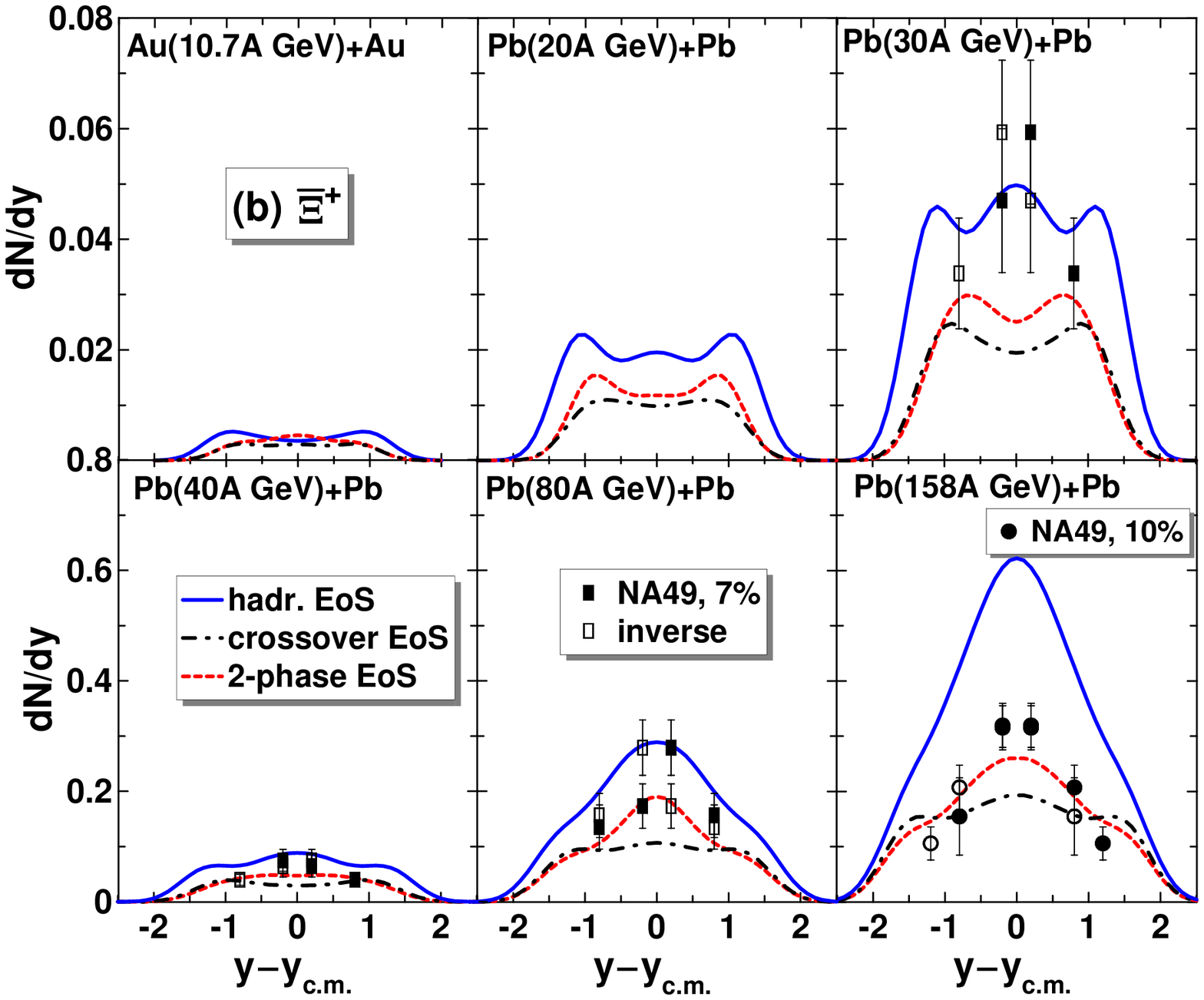}
 \caption{(Color online)
Rapidity spectra of $\Xi^-$ [(a) block of panels] and $\bar{\Xi}^+$  [(b) block of panels] hyperons
from central 
collisions of Au+Au ($b=$ 2 fm) and Pb+Pb ($b=$ 2.4 fm).   
Experimental data are taken for central  Pb+Pb collisions
\cite{Alt:2008qm}  at SPS energies 
with centrality 7\% at  20$A$--80$A$ GeV  and centrality 10\% at 158$A$ GeV. 
Results for Pb+Pb collisions
with $b=$ 4.6 fm are also displayed for $\Xi^-$'s at 158$A$ GeV.  
} 
\label{fig_Xi}
\end{figure*}
\begin{figure*}[htb]
\includegraphics[width=11.9cm]{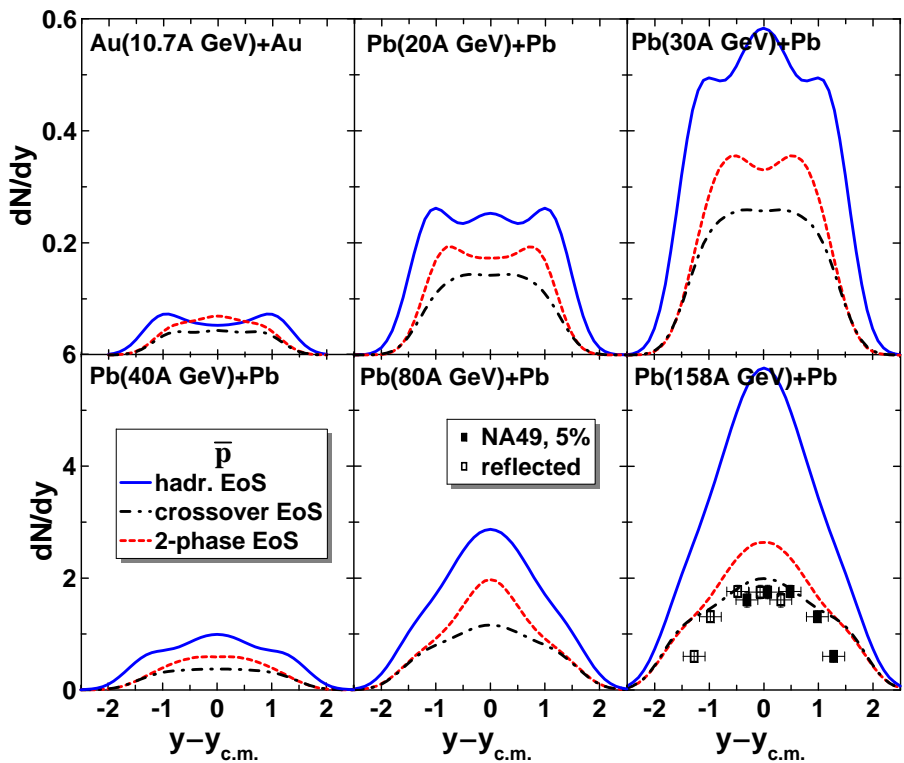}
 \caption{(Color online)
Rapidity spectra of anti-protons
from central 
collisions of Au+Au ($b=$ 2 fm) and Pb+Pb ($b=$ 2.4 fm).   
Experimental data for central  Pb+Pb collisions at 158$A$ GeV
with centrality 5\% are taken from Ref. \cite{Anticic:2010mp}. 
} 
\label{fig_ap}
\end{figure*}
\begin{figure}[htb]
\includegraphics[width=7.9cm]{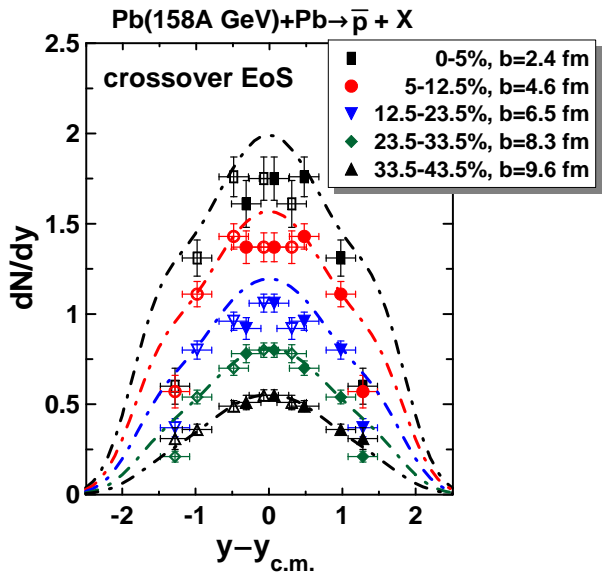}
 \caption{(Color online)
Rapidity spectra of anti-protons at various centralities
from  
collisions Pb+Pb at $E_{lab}=$ 158$A$ GeV.   
Experimental data are taken from Ref. 
\cite{Anticic:2010mp}. Curves present calculations with the crossover EoS 
at different impact parameters ($b$). 
\vspace*{-7mm}
} 
\label{fig_ap-158}
\end{figure}

\begin{figure*}[bth]
\includegraphics[width=13.95cm]{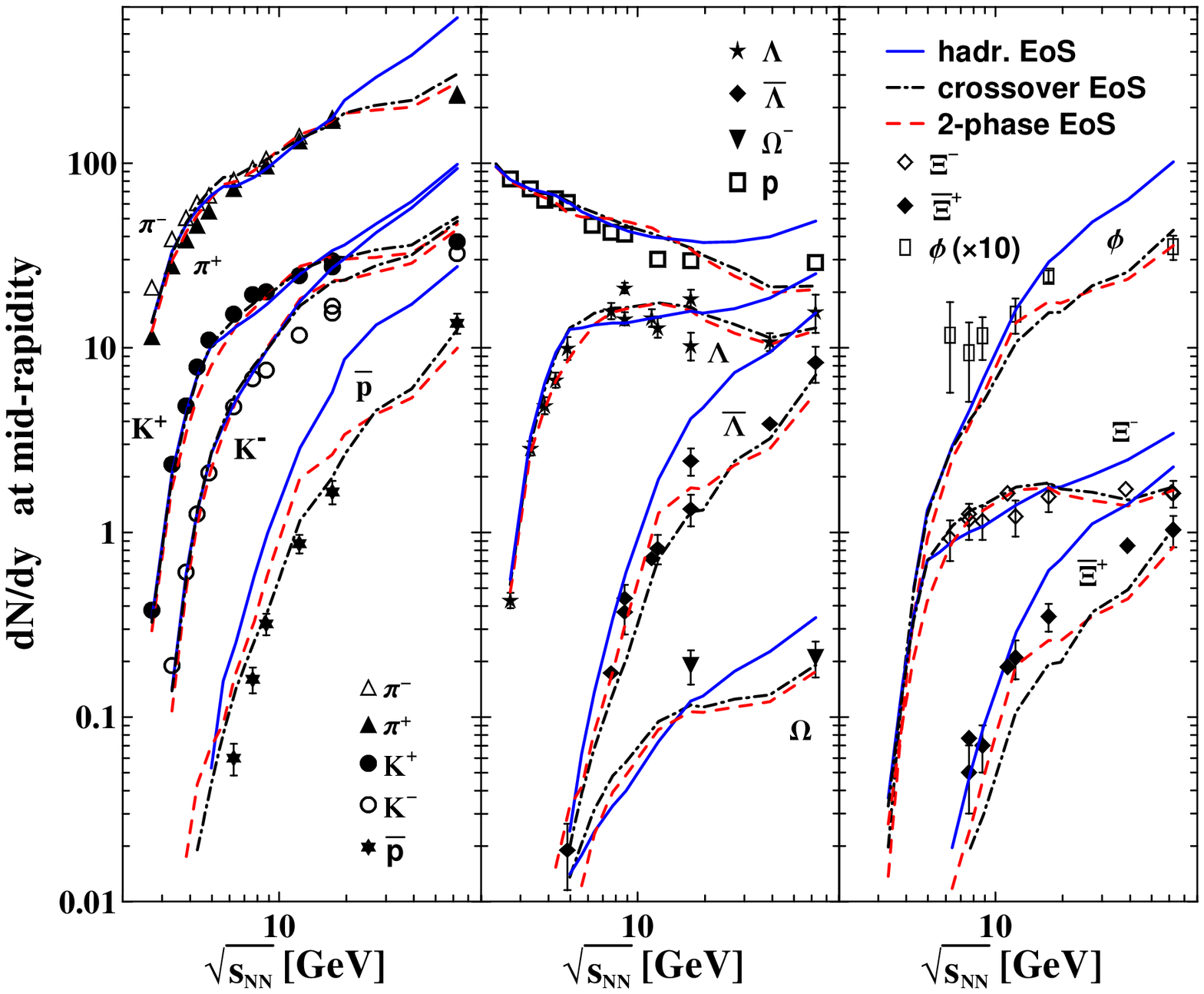}
 \caption{(Color online)
Rapidity densities $dN/dy$ at mid-rapidity of various produced particles 
as functions of the center-of-mass energy
 of colliding nuclei predicted
 by 3FD calculations with  three considered EoS's. 
Experimental data are from compilation of Ref. \cite{Andronic:2005yp}
complemented by recent data from STAR collaboration \cite{Zhu:2012ph} and  
latest update of the  
compilation of NA49 numerical results \cite{Compilation-NA49,Blume:2011sb}.
}  
\label{fig-cm-all}
\end{figure*}

In Fig. \ref{fig_K}, rapidity distributions of positive and negative kaons from 
central collisions are presented. In fact, the model provides  distributions of
kaons, $K$, and antikaons, $\bar{K}$, which are sums of yields of $K^+ + K^0$ and $K^- + \bar{K}^0$, 
respectively. The distributions of $K^+$ and $K^-$ are then approximately 
associated with distributions of $K$ and  $\bar{K}$, respectively, divided by two.  
This approximation induces a certain error especially at low incident energies, 
where  yields of $K^+$ and $K^0$ (and, correspondingly, $K^-$ and $\bar{K}^0$) 
noticeably differ.

As seen from Fig. \ref{fig_K}, the $K^-$ production is distinctly overestimated 
in all of the considered scenarios. However, it is early to conclude that all the scenarios
fail. This overestimation can be a consequence of the fact that the extrapolation of 
experimental data (the exponential one) beyond the measured points 
(thin solid lines in Fig. \ref{fig_K-mt})
essentially differs from results of calculations at high transverse masses ($m_T$), 
while in the experimentally measured low-$m_T$ region these are quite similar.   
In fact, determination of the rapidity density requires integration over $m_T$ 
up to infinity. Therefore, the experimental result can be different from the 
model ones because of their difference in the $m_T$ region, where experimental data 
are absent. Similar situation takes place for the $K^+$ yields. At lower SPS energies 
the  experimental extrapolation to high  $m_T$ reasonably agree with model results and 
hence the same reasonable agreement is observed in rapidity distributions. At higher SPS energies
(80$A$ and 158$A$ GeV) the experimental (exponential) extrapolation is again distinctly lower than  
the model predictions. Therefore, the model rapidity distributions overestimate the 
experimental ones. 
It is worthwhile to notice that the calculated shapes of the $m_T$ spectra of kaons are somewhat similar
to that of pion spectra, i.e. they are slightly concave in the logarithmic scale rather than linear, 
as it was assumed in the experimental extrapolation. 
Measurements of high transverse momenta later performed by NA49 collaboration 
at 158$A$ GeV \cite{Alt:2007cd}  showed 
that the exponential extrapolation of the low-$m_T$ data indeed underestimates the high-$m_T$ data, 
see (c) panel of Fig. \ref{fig_K-mt}. At the same time the 3FD predictions overestimate these 
high-$m_T$ data. The latter fact is expected, because even abundant hadronic species become rare 
species at high momenta. Therefore, their treatment on the basis of grand canonical ensemble
(which is the case in the 3FD model) results in overestimation of their yield.

In view of the above possible explanation of the difference  between calculated and experimental 
spectra it is early to conclude that the 3FD model fails to reproduce the 
$K^+$ and $K^-$  rapidity distributions. Again different EoS-scenarios agree (or disagree) with 
the available data approximately to the same extent. An important feature of the $K^+$ distributions
is that all scenarios underestimate mid-rapidity values of the experimental data. This 
results in a failure to reproduce the experimental ``horn'' in the $K^+/\pi^+$ ratio \cite{Alt:2007aa}, 
see sect.  \ref{Ratios}.

In Fig. \ref{fig_phi} rapidity distributions of $\phi$ mesons  from 
central collisions are presented. Here the situation is controversal. 
At high SPS energies, predictions of the hadronic EoS look certainly preferable. 
At lower  SPS energies 
{
and the top AGS energy, 
}
predictions of different scenarios are rather similar and 
all of them considerably underestimate the experimental data at the mid-rapidity region. 
A reason of this is still unclear.

Let us turn to baryonic distributions. 
Since proton distributions have been already analyzed in detail in Refs. 
\cite{Ivanov:2012bh,3FD-2012-stopping}, let us directly proceed to strange baryons. 
In Figs. \ref{fig_L} and \ref{fig_Xi} rapidity distributions of $\Lambda$, $\bar{\Lambda}$,   
$\Xi^-$  and $\bar{\Xi}^+$
hyperons from central collisions are presented.
Because of the low centrality at 158$A$ GeV (10\%), results for Pb+Pb collisions
with $b=$ 4.6 fm are also displayed for $\Lambda$ and $\Xi^-$. 
For corresponding antiparticles $\bar{\Lambda}$ and $\bar{\Xi}^+$ 
the results with $b=$ 2.4 fm and $b=$ 4.6 fm are fairly close to each other. 
Therefore, calculations with $b=$ 4.6 fm are not presented. 
As seen, data on $\Lambda$ and $\Xi^-$ hyperons are reasonably reproduced by all of 
considered EoS's.  
{
Apparent preference of the hadronic scenario in reproducing the $\Xi^-$ data 
cannot be considered as an argument in favor of this scenario because the extent of the 
overestimation by the deconfinement-transition scenarios is just within the accuracy of 
the grand-canonical approach. 
At the same time, for  anti-hyperons $\bar{\Lambda}$ and $\bar{\Xi}^+$ 
the hadronic scenario evidently fails: it considerably overestimates the data, 
especially at the top SPS energy. 
}
The case of $\bar{\Xi}^+$ at 30$A$ GeV is not spectacular. There the data have 
too large error bars. For $\Lambda$ and  $\bar{\Lambda}$ the crossover EoS looks 
certainly preferable. 
At the same time it is hardly possible to expect from the 3FD model 
with 2-phase and crossover EoS's a better agreement 
with data on such rare probes as $\Xi^-$ and $\bar{\Xi}^+$ because the  
model is based on grand-canonical statistics and therefore requires 
``large'' multiplicities to be valid.

Figure \ref{fig_ap} confirms that the hadronic scenario strongly overestimates experimental data 
on antibaryons, in the present case, on antiprotons. At the same time, the crossover scenario 
almost perfectly reproduces the antiproton rapidity distribution even at various centralities, see Fig.  
\ref{fig_ap-158}. Correspondence between the centrality related to a data set 
and a mean value of  the impact parameter used in the calculation is taken from Ref. \cite{Alt:2003ab}.

In all the blocks of figures I kept the results for Au+Au collisions at 10$A$ GeV, even if there 
are no data for this reaction. This done because the  deconfinement transition in the presently 
studied scenarios of nuclear collisions starts around this incident energy, 
see Refs. \cite{Ivanov:2012bh,3FD-2012-stopping}. Therefore, it is important to observe evolution 
of particle distributions beginning from this energy.

\begin{figure}[tbh]
\includegraphics[width=5.0cm]{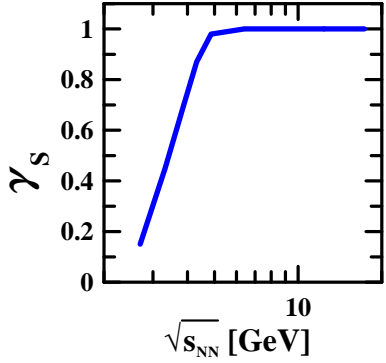}
 \caption{(Color online)
Strangeness suppression factor  as a function of the center-of-mass energy
 of colliding nuclei. 
}  
\label{fig-gammaS}
\end{figure}
\begin{figure}[thb]
\includegraphics[width=6.7cm]{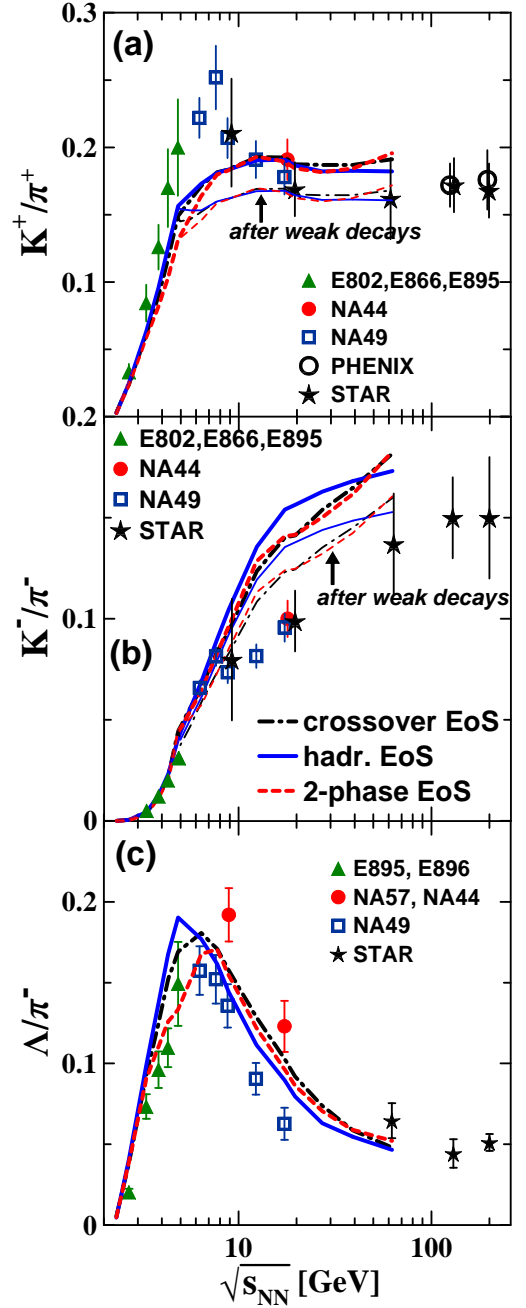}
 \caption{(Color online)
The energy dependence of ratios of the rapidity densities $dN/dy$ at mid-rapidity: 
(a) $K^+$ to $\pi^+$,
(b) $K^-$ to $\pi^-$,  and (c) $\Lambda$ to $\pi^-$. 
Compilation of experimental data is from Ref. 
\cite{Blume:2011sb}  
complemented by recent STAR data \cite{Zhu:2012ph,Aggarwal:2010ig}.
Calculated ratios with pions including contribution from weak decays 
are also displayed by thin lines. The latter ratios are relevant to STAR and PHENIX data.  
} 
\label{fig_ratios}
\end{figure}

\section{Mid-rapidity Densities and Their Ratios}
\label{Ratios}

Fig. \ref{fig-cm-all} summarizes and extends the discussion of the previous section. 
It displays 
excitation functions of the mid-rapidity  values 
of the rapidity spectrum of various produced particles but in a wider range of 
incident energies $\sqrt{s_{NN}}=$ 2.7--62.4 GeV.  
Results for the top 
calculated energy of $\sqrt{s_{NN}}=$ 62.4 GeV should be taken with care,  
because they are not quite accurate, as an accurate computation requires 
unreasonably high memory and CPU time. 

The strangeness production at low incident
energies is overestimated within the 3FD model. This is 
the consequence of the fact that the 3FD model
is based on the grand canonical ensemble. This shortcoming
can be easily cured by introduction of a phenomenological factor
$\gamma_S$  \cite{Koch:1986ud} which takes into 
account an additional strangeness 
suppression due to constraints of canonical ensemble. 
The mid-rapidity densities of single-strange particles 
($K^\pm$, $\Lambda$ and $\bar{\Lambda}$), displayed in Fig. \ref{fig-cm-all}, 
are multiplied by $\gamma_S$ factor, and multi-strange particles
($\phi$, $\Xi^-$, $\bar{\Xi}^+$ and $\Omega$), by  $(\gamma_S)^{n_s}$ factor, 
where $n_s$ is the number of valence strange quarks contained in this particle. 

The excitation function of the $\gamma_S$ factor is presented in 
Fig. \ref{fig-gammaS}, which is of course applicable only to central 
collisions of considered nuclei. 
On average, there is no
need for additional strangeness suppression at $\sqrt{s_{NN}}>$ 5 GeV, 
though the reproduction of data on some strange species is still far from 
being perfect. The hadronic scenario looks worst in this respect. 
At 5 GeV  $\le \sqrt{s_{NN}}\le$ 17 GeV, i.e. in the range of the SPS energy, 
it is impossible to find a unique value of $\gamma_S$ for this scenario to simultaneously 
improve reproduction of data on all of the strange hadrons. 
At $\sqrt{s_{NN}}\ge$ 17 GeV the hadronic scenario again requires an 
additional strangeness suppression. However, introduction of such a suppression
at high incident energies is physically unjustified, because multiplicities 
of strange hadrons are already quite high for applicability of 
the grand canonical ensemble.


Fig. \ref{fig-cm-all} confirms that the hadronic scenario considerably overestimates 
mid-rapidity densities of antibaryons (strange and nonstrange) predicted by 
EoS's with a  deconfinement transition  and (as a rule) those deduced form experiment, 
starting already from lower SPS energies, i.e. $\approx$20$A$ GeV. 
At energies above the top SPS one, the mid-rapidity densities predicted 
by the hadronic scenario considerably exceed the data on all species. 
Notice that agreement of hadronic-scenario predictions with data on the major part of species 
in the SPS energy range was achieved at the expense of considerable enhancement 
the inter-fluid friction in the hadronic phase \cite{3FD-2012-stopping,3FD} 
as compared with its microscopic estimate of Ref. \cite{Sat90}.
Precisely this enhancement makes bad job at energies above the SPS.

The advantage of  deconfinement-transition scenarios is that they do not require any modification 
of the microscopic friction in the hadronic phase. 
They reasonably agree with all the data with the exception of negative kaons and $\phi$-mesons  
in the SPS energy range. A possible reason of disagreement with negative-kaon data is 
inaccuracy of the experimental extrapolation of the transverse-mass spectra, 
discussed in the previous section. As for $\phi$-mesons, the reason of this is still unclear. 
Rare probes, such as $\bar{\Xi}^+$ and $\Omega^-$, are also poorly reproduced, that is a 
consequence of inapplicability of the grand canonical ensemble for the treatment of such   
rare species.

In Fig. \ref{fig_ratios}, ratios of  various hadronic mid-rapidity densities are presented 
as functions of the incident energy. 
Calculated ratios with pions including contribution from weak decays of strange hadrons
are also displayed by the corresponding thin lines.    
These ratios are relevant to STAR and PHENIX data 
\cite{Zhu:2012ph,Aggarwal:2010ig}. 
All of the considered scenarios do not reproduce the horn anomaly in 
the $K^+/\pi^+$ ratio \cite{Alt:2007aa}
around $E_{lab}=$ 30$A$ GeV. As it was demonstrated in Fig. \ref{fig_K}, the reason is 
that all scenarios underestimate mid-rapidity values of $K^+$ as compared to the experimental data, 
while its rapidity distributions at peripheral rapidities look almost perfect. 
As it was found, the value of this ratio is quite insensitive to variations of the 
freeze-out parameter $\varepsilon_{\scr frz}$
\cite{3FD-2012-stopping,3FD}, because the yields of $K^+$ and $\pi^+$ increase or 
decrease proportionally with the change of $\varepsilon_{\scr frz}$.

{
Authors of Ref. \cite{Andronic:2008gu} have noticed that within the statistical model 
the peak the $K^+/\pi^+$ ratio  can be better reproduced if
the contribution of the $f_0$(600) resonance, i.e. the $\sigma$ meson, is taken into account. 
However, authors of Ref. \cite{Satarov:2009zx} concluded that the $\sigma$ meson
does not solve the problem. In the present 3FD calculations the $\sigma$ meson is included.  
 Moreover, the $\sigma$ spectral 
function is taken into account accordingly to parametrization of Ref. \cite{Wolf:1997iba}.
}

A possible way to 
improve the reproduction of the horn anomaly 
is to accept 
different freeze-out conditions, i.e. energy densities $\varepsilon_{\scr frz}$, for $K^+$ and $\pi^+$ 
in this narrow\footnote{In a wide energy range, the universal freeze-out for $K^+$ and $\pi^+$
works faily well and there are no reasons to change it.} 
energy region near $E_{lab}=$ 30$A$ GeV. However, physical reasons for such a 
radical solution of the horn-anomaly problem are absent. Another possibility 
consists in noticeable difference in $K^+$ and $K^0$ densities in this energy range. 
I would like to remind that the 3FD model does not distinguish isotopic states of hadrons 
and hence the yield of $K^+$ is determined as half of the total 
$K$ yield consisting of the sum of $K^+$ and $K^0$ contributions. 
Similarly, the $\pi^+$ yield is determined as one third of the total pion one. 
However, a physical reason for such an isotopic asymmetry in this narrow energy range 
is not seen. Moreover, such an asymmetry would be opposite to a natural isotopic asymmetry 
at which the $K^0$ yield exceeds the $K^+$ one. This natural asymmetry results from  
initial isotopic asymmetry of colliding nuclei.

A possible reason for overestimation of $K^-/\pi^-$
ratio, see (b) panel of Fig. \ref{fig_ratios}, as compared with the data 
has been already mentioned in the previous section. It can result from difference of 
the extrapolation of transverse-mass experimental data 
(exponential extrapolation) beyond the measured points 
(thin solid lines in Fig. \ref{fig_K-mt}) 
from results of calculations at high transverse masses ($m_T$).   
Therefore, it would be desirable to directly compare the transverse-momentum 
spectra predicted by the model with  experimental ones at high 
transverse momenta. 
The $\Lambda/\pi^-$ ratio, see (c) panel of Fig. \ref{fig_ratios}, 
is quite satisfactorily reproduced by all of the scenarios.

\section{Summary}

Results on rapidity distributions of particles produced in 
relativistic heavy-ion collisions in the energy range 2.7 GeV $\le \sqrt{s_{NN}} \le$
62.4 GeV  are presented.  
These simulations were performed within the three-fluid model
\cite{3FD} employing three different equations of state: a purely hadronic EoS   
\cite{gasEOS} and two versions of the EoS involving the deconfinement 
 transition \cite{Toneev06}. These two versions are an EoS with the first-order phase transition
and that with a smooth crossover transition. 
Details of these calculations are described in the first paper of this series 
\cite{3FD-2012-stopping} dedicated to analysis of the baryon stopping.

If was found that the model within all of the considered scenarios reproduces approximately to the same extent 
the available data on almost all (with some exceptions) hadronic species 
in the AGS--SPS energy range, i.e. at 2.7 GeV $\le \sqrt{s_{NN}} \le$ 17.4 GeV. 
In the case of the hadronic EoS this agreement is 
 achieved at the expense of noticeable enhancement of 
the inter-fluid friction in the hadronic phase \cite{3FD-2012-stopping,3FD} 
as compared with its microscopic estimate 
of Ref. \cite{Sat90}.
{
Reproduction of the data is better for abundant species, as it should be 
within grand-canonical statistics.
The above mentioned exceptions are 
\begin{itemize}
	\item 
At lower SPS energies and the top AGS energy, predictions of all scenarios are rather similar and 
all of them considerably underestimate the experimental $\phi$-meson data. 
Moreover, at the top SPS energy, the hadronic EoS unexpectedly 
gives the best description of the $\phi$-meson data. 
The reason of this is unclear. 
	\item 
The 
strangeness enhancement in the incident energy range near 30$A$ GeV,  
i.e. the ``horn'' anomaly in the $K^+/\pi^+$ ratio \cite{Alt:2007aa}, 
is not reproduced in any scenario. 
	\item 
The $K^-$ production is distinctly overestimated 
in all of the considered scenarios. 
This overestimation can be a consequence of the extrapolation of 
experimental data (the exponential one) beyond the measured points that 
essentially differs from results of calculations at high transverse masses, 
while in the experimentally measured low-$m_T$ region these are quite similar. 
	\item 
Data of rare probes (like $\Xi^-$  and $\bar{\Xi}^+$ hyperons) are poorly reproduced 
apparently because of the grand-canonical statistics used in the 3FD model. 
	\item 
The last exception concerns only the hadronic EoS. 
The hadronic scenario considerably overestimates experimental 
rapidity densities of antibaryons (both strange and nonstrange), 
starting already from lower SPS energies, i.e. $\approx$ 20$A$ GeV.   
At the same time, the deconfinement-transition scenarios successfully (to a various extent) reproduce 
these data. 
\end{itemize}
The failure with the $\phi$-meson data and the ``horn'' anomaly requires further investigation. 
The proposed reason of the overestimation of the $K^-$ production also needs to be checked.  
}

At energies above the top SPS one, i.e. at $\sqrt{s_{NN}}>$ 17.4 GeV, the mid-rapidity densities predicted 
by the hadronic scenario considerably exceed the available RHIC data on all the species. 
The  deconfinement-transition scenarios reasonably agree with all the data.

The 3FD model 
deals with three different fluids    \cite{3FD,3FD-2012-stopping}.  
These are two baryon-rich fluids 
initially associated with constituent nucleons of the projectile
and target nuclei. These fluids
are either spatially separated or 
unified at the freeze-out stage.  
In addition, newly produced particles,
populating the mid-rapidity region, are associated with a 
baryon-free (``fireball'') fluid
which remains undissolved in baryonic fluids till the freeze-out. 
A certain formation time $\tau$ is allowed for the fireball fluid, during
which the matter of the fluid propagates without interactions. 
The formation time is  associated with a finite time of string formation.
The main difference concerning this baryon-free fluid in considered alternative 
scenarios consists in different formation times: $\tau = 2 \;\mbox{fm/c}$ 
for the hadronic scenario and $\tau = 0.17 \;\mbox{fm/c}$ for scenarios involving 
the deconfinement transition  \cite{3FD-2012-stopping}.

As it is seen from simulations, the main contribution to baryon and meson yields 
comes from baryon-rich fluids within the deconfinement-transition scenarios. 
This contribution slowly decreases  
with the incident energy rise. For instance, at 
$\sqrt{s_{NN}}=$ 39 GeV approximately half of pions are produced from the 
baryon-free fluid within the deconfinement-transition scenarios. 
For all other particles (i.e. baryons and mesons) and lower energies this 
fraction is essentially lower. 
At the same time the fraction of half for pions from the 
baryon-free fluid is achieved already at $\sqrt{s_{NN}}\simeq$ 9 GeV 
(i.e. $E_{lab}\simeq$ 40$A$ GeV) within the hadronic scenario. 
This is one of the reasons why $\tau$ was chosen so large in the 
hadronic scenario. 
Large formation time prevents absorption of the baryon-free matter by 
the baryonic fluids. 
Without this large contribution of the baryon-free fluid
it is impossible to reproduce mesonic yields at SPS energies within the hadronic scenario. 
However, this strongly developed baryon-free fluid makes bad job for 
antibaryons in the case of the hadronic EoS. The reason is that 
antibaryons are dominantly 
produced from the baryon-free fluid even at lower incident energies
within all of the considered scenarios. 
Their yields in the hadronic scenario strongly overestimate experimental data. 
At $\sqrt{s_{NN}}>$ 17.4 GeV the overdeveloped baryon-free fluid makes already bad job for
for all produced particles, resulting in considerable overestimation of 
the available RHIC data on all the species within the hadronic scenario.

All the above speculations are formulated in terms of the 3FD model. However, 
they hold true in general physical terms, if baryon-rich fluids are understood as a matter 
formed of leading particles, predominantly (but not solely) occupying peripheral rapidities, 
and the baryon-free fluid is associated with newly produced particles, 
predominantly (but not solely) occupying  mid-rapidity region. At large baryon stopping, 
the difference between these groups of particles become less pronounced.

All the above results certainly indicate in favor of the deconfinement-transition scenarios, 
though they also are not perfect in reproduction of the data ($\phi$-meson, the ``horn'' anomaly).  
It is difficult to judge on a preference of the 2-phase or crossover EoS's. 
The first-order transition in the 2-phase EoS is very abrupt. On the contrary, 
the crossover transition constructed in Ref. \cite{Toneev06}
is very smooth  \cite{Ivanov:2012bh,3FD-2012-stopping}.  
In this respect, this version of the crossover EoS certainly contradicts results of the 
lattice QCD calculations, where a fast crossover, at least at zero chemical potential, 
was found \cite{Aoki:2006we}. 
Therefore, a true EoS is somewhere in between  the crossover and 2-phase EoS's 
of Ref. \cite{Toneev06}. 

\vspace*{3mm} {\bf Acknowledgements} \vspace*{2mm}

I am grateful to A.S. Khvorostukhin, V.V. Skokov,  and V.D. Toneev for providing 
me with the tabulated 2-phase and crossover EoS's. 
The calculations were performed at the computer cluster of GSI (Darmstadt). 
This work was supported by The Foundation for Internet Development (Moscow)
and also partially supported  by  the grant NS-215.2012.2.

\end{document}